\documentclass[aps,prl,superscriptaddress,amsfonts,amsmath,amssymb,showpacs,floatfix,reprint,longbibliography]{revtex4-2}

\usepackage{url}
\usepackage{bm}
\usepackage{graphicx}
\usepackage{epsfig}
\usepackage{amsmath}
\usepackage{amstext}
\usepackage{amssymb}
\usepackage{amsfonts}
\usepackage{amsbsy}
\usepackage{verbatim}
\usepackage{xcolor}
\usepackage{hyperref}
\hypersetup{colorlinks=true, urlcolor=blue, linkcolor=blue, citecolor=blue}
\usepackage{multirow}
\usepackage{floatrow}
\usepackage{float}
\usepackage{gensymb}
\usepackage{textcomp}
\usepackage{tabularx}
\usepackage{enumitem}
\usepackage[version=4]{mhchem}
\usepackage{afterpage}
\usepackage{pbox}
\usepackage{makecell}
\usepackage{array,booktabs}
\usepackage{dsfont}
\usepackage{siunitx}
\usepackage[utf8]{inputenc}
\usepackage{braket}
\usepackage[normalem]{ulem}

\graphicspath{ {FiguresMain/} }

\begin{document}

\title{Structure, heat capacity and Raman spectra of mm-sized Ba$_{2}$MgWO$_{6}$ single crystals synthesized by BaCl$_{2}$-MgCl$_{2}$ flux method}

\author{Jana P\'{a}sztorov\'{a}*}
\email{pasztorovaj@gmail.com}
\affiliation{Laboratory for Quantum Magnetism, Institute of Physics, \'Ecole Polytechnique F\'ed\'erale de Lausanne, CH-1015 Lausanne, Switzerland}

\author{Wen Hua Bi}
\affiliation{Crystal Growth Facility, \'Ecole Polytechnique F\'ed\'erale de Lausanne, CH-1015 Lausanne, Switzerland}

\author{Richard Gaal}
\affiliation{Laboratory for Quantum Magnetism, Institute of Physics, \'Ecole Polytechnique F\'ed\'erale de Lausanne, CH-1015 Lausanne, Switzerland}

\author{Karl Kr\"amer}
\affiliation{Department of Chemistry, Biochemistry and Pharmacy, University of Bern, CH-3012 Bern, Switzerland}

\author{Ivica {\v Z}ivkovi{\' c}}
\affiliation{Laboratory for Quantum Magnetism, Institute of Physics, \'Ecole Polytechnique F\'ed\'erale de Lausanne, CH-1015 Lausanne, Switzerland}

\author{Henrik M. Rønnow}
\affiliation{Laboratory for Quantum Magnetism, Institute of Physics, \'Ecole Polytechnique F\'ed\'erale de Lausanne, CH-1015 Lausanne, Switzerland}

\date{\today}

\begin{abstract}

{We present a new method of Ba$_{2}$MgWO$_{6}$ single crystal synthesis that allows to grow larger crystals using BaCl$_{2}$ and MgCl$_{2}$ flux. Difficulties to grow single crystal of a size suitable for macroscopic material property measurements caused the majority of characterisation being published on polycrystalline samples. Single crystal diffraction and energy dispersive X-ray analysis confirmed high quality of synthesised samples. Heat capacity measurements from 300~K to 2~K do not show any transitions. However, Raman spectra measured down to 77~K contain additional peaks at all temperatures probed, which is in a contrast with only 4 Raman active modes expected from the reducible representation. This calls for a more detailed study of potential symmetry breaking that could also influence the electronic properties of the material.}
    
\end{abstract}

\maketitle

\section{Introduction}

Perovskites $ABO_{3}$, standardly having $Pm\bar{3}m$ symmetry, were in a scope of scientific interest for a long time due to a vast variety of electronic, magnetic and optical properties \cite{ob:perovskites,jr:shirane}. In case of double perovskites $A_{2}BB'O_{6}$, the previous $B$ site is now occupied by two different types of cations, which naturally increases the variety of properties \cite{jr:P_DP} and reduces the symmetry into $Fm\bar{3}m$. As the typical oxidation state of tungsten is 6+, it is an ideal candidate for $B'$ site occupancy in $A_{2}BB'O_{6}$, creating a bond with 6 oxygen atoms in the octahedral fashion. The early studies on A$_{2}$BWO$_{6}$ \cite{jr:1951_pow,jr:diel_Ram_1974,jr:pow_A2BXO6} were establishing structural properties and octahedral distortion present in polycrystalline samples mostly. This is particularly important when assessing electric properties of a material, as the octahedral tilting can break centro-symmetry which is critical for ferroelectricity. Another field of interest in tungstates comes from their potential as ceramic materials for microwave applications \cite{jr:diel_Ram1}. Theoretical calculations of Ba$_{2}$MgWO$_{6}$ phonon spectra were also made utilising density functional theory (DFT) \cite{jr:diel_Ram1} and lattice dynamic calculations (LDC) \cite{jr:LDC}. To the best of our knowledge, there is only one published paper about the single crystal growth of Ba$_{2}$MgWO$_{6}$ in which molten K$_{2}$CO$_{3}$ flux was used \cite{jr:WO6_pow}. To date no single crystal measurements of materials properties have been reported.

	 In this paper, we will present growth of Ba$_{2}$MgWO$_{6}$ single crystals that are larger in size and were produced in a new way using oxides - BaO, MgO WO$_{3}$ - and BaCl$_{2}$, MgCl$_{2}$ as a flux in Pt tubes - eliminating the risk of potassium and aluminium  contamination. The idea was based on successful growth of Ba$_{2}$MgReO$_{6}$ single crystals \cite{jr:hirai,jr:Ba2MgReO6_jana}. Moreover, the W$^{6+}$ ion size in non-magnetic  Ba$_{2}$MgWO$_{6}$ is similar to Re$^{6+}$ ionic radius. Finally, we report specific heat and Raman spectra measured on single crystals.

\section{Experimental Procedure}
	 
	 \subsection{Materials}
	 The following chemicals were used for the synthesis: BaO (Sigma Aldrich, 99.99$\%$), MgO (Apollo Scientific, 99.99$\%$), WO$_{3}$ (Fluka, 99.9$\%$), and BaCl$_{2}$ (ABCR Swiss AG, ultra dry, 99.998$\%$), MgCl$_{2}$ (ABCR Swiss AG, ultra dry, 99.99$\%$) as a flux.
	 
	  \subsection{Single crystal growth}
	  
	 BaO, MgO and WO$_{3}$ were mixed in stoichiometric ratios and ground thoroughly in an agate mortar in an Ar glove box to avoid moisture. The flux composed of 71 wt$\%$ BaCl$_{2}$ and 29 wt$\%$ MgCl$_{2}$ was added and mixed thoroughly. The mixture with the total mass of 4~g was then placed in a platinum tube of 95 mm length and 10 mm diameter and sealed using the arc meting machine with the protective Ar atmosphere. The tube was heated to 1400 $^{\circ}$C where it was kept for 15 h and then slowly cooled to 950 $^{\circ}$C at a rate of 1.5 $^{\circ}$C/h, followed by furnace cooling to room temperature. 
	 The excess flux was removed by washing with distilled water in an ultrasonic bath. Crystals were isolated by the vacuum filtration.
	  
	  \subsection{Scanning electron microscopy}
	  
	  The chemical composition of the single-crystalline Ba$_{2}$MgWO$_{6}$ samples was verified by energy dispersive X-ray spectroscopy in a scanning electron microscope (SEM-EDX, Zeiss GeminiSEM 300 with Oxford Inst. EDX detector) in high-vacuum mode. EDX spectroscopy confirmed the presence of Ba, Mg, W and their respective stoichiometric ratios within the elemental detection limits of the instrument.

	   \subsection{Single crystal X-ray diffraction}
	   
	   The quality of the single crystals was checked using a Rigaku Synergy-I XtaLAB X-ray diffractometer, equipped with a Mo micro-focusing source ($\lambda$K$_{\alpha}$~=~0.71073~\AA) and a HyPix-3000 Hybrid Pixel Array detector (Bantam). According to the results of data reduction with \textit{CrysAlisPro} program, all diffraction spots can be indexed by the F-centered cubic unit cell. The refinement on collected data using \textit{OLEX2} and \textit{ShelXL} software.
	   
	   \subsection{Heat capacity}
	   
	   Specific heat was measured by Quantum Design Physical Property Measurement System utilising the relaxation method.

	   \subsection{Raman Spectroscopy}

	    Raman spectra of Ba$_{2}$MgWO$_{6}$ single crystals was measured on a Horiba HR-800 spectrometer equipped with a 532 nm laser, a 600 l/mm grating and a Symphony II nitrogen cooled CCD detector. The objective used was a 50x long working distance lens from Olympus. The sample was mounted on the cold finger of a Conti continuous flow cryostat. Spectra were taken every 50 K, from 300~K down to 100~K  and at 77~K. Peaks were identified and fitted with a Pearson VII line shape.

\section{Results and discussion}

    \subsection{Preparation, element and structural analysis}

\begin{figure}[h]
        \includegraphics[width=0.4\columnwidth,trim= 3cm 1.7cm 1cm 0,clip]{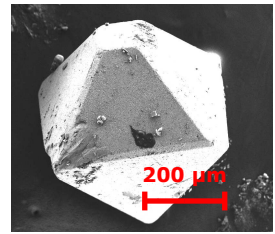}
     	\centering
     	\caption{The SEM photograph of a typical Ba$_{2}$MgWO$_{6}$ single crystal with truncated octahedral shape.} 
     	\label{fig:SC}
    \end{figure}

The grown crystals have average size of 0.4~mm and typical morphology of truncated octahedron as can be seen in Fig.~\ref{fig:SC}. The maximum size was 0.6~mm (used for heat capacity measurement, Fig.~\ref{fig:HC}). The growth had a high yield of crystals - for total mass of 4~g of starting materials and flux, there was single crystal yield of 0.5~g. Crystals have semi-transparent light brown colour.

\begin{figure}[h]
    	\includegraphics[width=0.8\columnwidth]{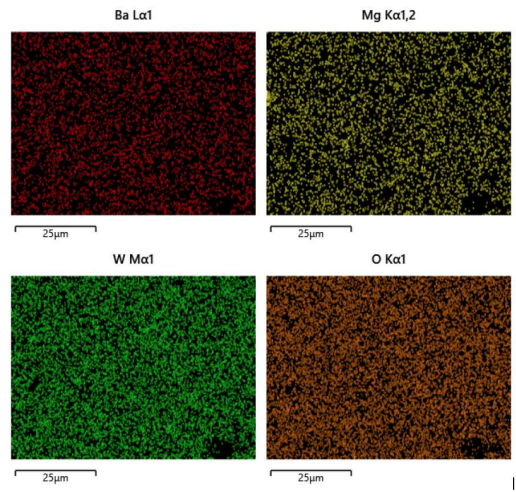}
     	\centering
     	\caption{Uniform distribution of Ba, W, Mg and O elements measured on Ba$_{2}$MgWO$_{6}$ single crystal.} 
     	\label{fig:EDX_elem}
    \end{figure}

\begin{figure}[h!]
     	\includegraphics[width=0.8\columnwidth]{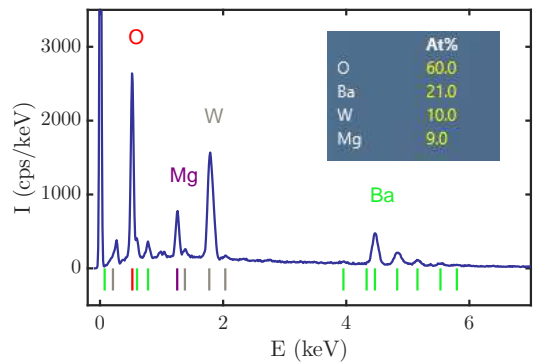}
     	\centering
     	\caption{EDX analysis on Ba$_{2}$MgWO$_{6}$ single crystal. Vertical lines underneath the spectra indicate x-ray peaks of individual elements. The table shows atomic percentages of each element, which was consistent for measurements on other crystals and facets. The fit of peak position and intensity was performed automatically by Zeiss GeminiSEM software.} 
     	\label{fig:EDX}
    \end{figure}

EDX spectra and element distribution maps were recorded from several selected crystals and from different facets on each of these single crystals. The elemental mapping (Fig.~\ref{fig:EDX_elem}) demonstrates that the Ba, Mg, W and O atoms are uniformly distributed on the surface of tested single crystals within the detection limit of the technique. The molar ratio of Ba, Mg, W and O are determined to be very close to the stoichiometric Ba$_{2}$MgWO$_{6}$ sample (Fig.~\ref{fig:EDX}). We consider the error to be within $\pm$~2$\%$ limit.

\begin{figure}[h]
     	\includegraphics[width=0.9\columnwidth]{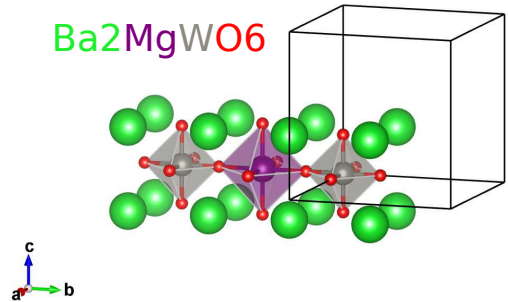}
     	\centering
     	\caption{The crystal structure of Ba$_{2}$MgWO$_{6}$ based on the refinement of single crystal X-ray data at 100~K. Cubic unit cell and W-O$_{6}$ and Mg-O$_{6}$ octahedrons were visualised in VESTA software.} 
     	\label{fig:octahed}
    \end{figure}

Single crystal X-ray diffraction was performed on a representative crystal at 300~K and 100~K. The lattice parameters for $Fm\bar{3}m$ space group, collection details and refinement information are given in Table~\ref{table:SC_XRD} and \ref{table:xyz_U}. Results are consistent with previous work \cite{jr:WO6_pow}. Ba$_{2}$MgWO$_{6}$ adopts the cubic double perovskite structure with corner shared octahedra of MgO$_{6}$ and WO$_{6}$ and Ba cations sitting in a space between, see Fig.~\ref{fig:octahed}. The bond lengths refined from the single crystal X-ray diffraction data at room temperature are following - Ba-O 2.867(12)~\AA, Mg-O 2.132(4)~\AA~and W-O 1.924(4)~\AA. The bonds become slightly shorter at 100~K - Ba-O 2.866(12)~\AA, Mg-O 2.119(3)~\AA~and W-O 1.931(3)~\AA. When comparing to the case of Ba$_{2}$MgReO$_{6}$, the interatomic distance of Ba-O atoms is rather similar. Larger difference occurs for Mg-O and $B'$-O distances - Mg-O is 2.082(5)~\AA~and Re-O 1.961(3)~\AA~\cite{jr:bramik} respectively.
    
    \begin{table}[h]
    \scriptsize
        \caption{Structure refinement details from single crystal X-ray diffraction on Ba$_{2}$MgWO$_{6}$, using the wavelength $\lambda$=~0.71073~\AA~and space group \textit{Fm$\bar{3}$m} at 100~K and 300~K.}
        \centering
        \begin{tabular}{l l l}
        \hline \\
        & 100~K & 300~K \\
        \hline \\ 
        Formula weight & 578.84 & 578.84 \\
        a(\AA) & 8.1010(2) & 8.1120(3) \\
        V(\AA$^{3}$) & 531.64(4) & 533.81(6) \\
        $\rho_{c}$ (g cm$^{-3}$) & 7.232 & 7.203 \\
        $\mu$ (mm$^{-1}$) & 36.310 & 36.163 \\
        F(000) & 984 & 984\\
        Crystal size (mm) & 0.04$\times$0.03$\times$0.02 & 0.01$\times$0.07$\times$0.03\\
        Reflections collected & 890 & 461\\
        Independent reflections & 95 & 64\\
        Goodness of fit on $F^{2}$ & 1.088 & 1.137\\
        R($F$) & 0.0105 & 0.0122\\
        R$_{w}$($F^2$) & 0.0251 & 0.0280\\
        Largest diff. peak/hole (e \AA$^{-3}$) & 0.698 and -0.555 & 1.090 and -0.642\\ [1ex]
        \hline
        \end{tabular}
        \label{table:SC_XRD}
    \end{table}

    \begin{table}[h]
    \scriptsize
        \caption{Atomic coordinates and equivalent isotropic displacement parameters for Ba$_{2}$MgWO$_{6}$, vales for single crystal X-ray structure determination collected at 100~K. $U_{eq}$ is defined as one third of the trace of the orthogonalized $U_{ij}$ tensor.}
        \centering
        \begin{tabular}{l l l l l}
        \hline\\
        Element & \textit{x} & \textit{y} & \textit{z} & $U_{ij}$ (\AA$^{2}$)\\
        \hline \\ 
        Ba & 0.25 & 0.25 & -0.25 & 0.003(1)\\
        Mg & 0 & 0 & 0 & 0.003(1)\\
        W & 0 & 0 & -0.5 & 0.002(1)\\
        O & 0 & 0 & -0.2384(4) & 0.004(1)\\
        \hline
        \end{tabular}
        \label{table:xyz_U}
    \end{table}

 We also examined the possibility of Mg/W site mixing which can occur in a case of double perovskites. For W site (4$b$), we confirm the occupancy factor equal to 1.00 and therefore Mg atoms can not sit on these positions. The occupancy factor for Mg site (4$a$) was 0.96, although excluding any option of a heavy element occupying it. We will refer to this result in the chapter about Raman spectra.

    \subsection{Heat Capacity}

\begin{figure}[h]
     	\includegraphics[width=0.7\columnwidth]{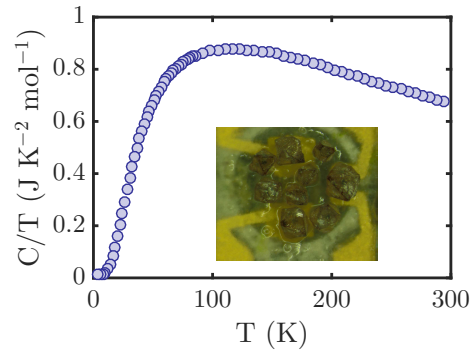}
     	\centering
     	\caption{Temperature dependence of specific heat divided by temperature C/T measured on Ba$_{2}$MgWO$_{6}$ single crystals showing no anomaly in data in temperature range from 300~K to 2~K. The inset shows single crystals on the PPMS puck platform with dimensions 3$\times$3~mm.} 
     	\label{fig:HC}
    \end{figure}

    Specific heat measurements used a sample puck with integrated heater and thermometer. Before mounting Ba$_{2}$MgWO$_{6}$ crystals on the platform, a background measurement was performed with small amount of the N apiezon grease. Subsequently, 9 octahedrally shaped crystals with the total mass of 3.6~mg were placed on the $\{$111$\}$ facet parallel to the puck. In Fig. \ref{fig:HC}, we show measured specific heat divided by temperature $C/T$ for temperature range from 300~K to 2~K in zero magnetic field. The specific heat follows the typical shape for phononic heat of an insulator and no indication of a phase transition induced anomaly was observed.
    
 	\subsection{Raman Spectroscopy}
 	
 	The allowed number of Raman active vibrational modes can be deduced from the site symmetry group analysis \cite{jr:bilbao_Ram} which for the face centered cubic Ba$_{2}$MgWO$_{6}$ perovskite structure with space group $Fm\bar{3}m$ has the following irreducible representation:

 	\begin{equation*}
        \Gamma = A_{1g} + E_{g} + F_{2u} + 2F_{2g} + 5F_{1u} + F_{1g},
    \end{equation*}

    where Raman and infrared active (IR) modes are:

    \begin{equation*}
        \begin{aligned}
         & \Gamma_{Ram} = A_{1g} + E_{g} + 2F_{2g}, \\
         & \Gamma_{IR} = 4F_{1u},
        \end{aligned}
    \end{equation*}

    $F_{1u}$ is the acoustic mode and $F_{1g}$, $F_{2u}$ are optically silent modes. $A$ denotes the nondegenerate mode, $E$ and $F$ denote the double and triple degeneracy modes, respectively. The symmetry analysis predicts only 4 Raman active modes that appear at different frequencies in the vibration spectrum.
    
    \begin{figure}[h]
     	\includegraphics[width=0.9\columnwidth]{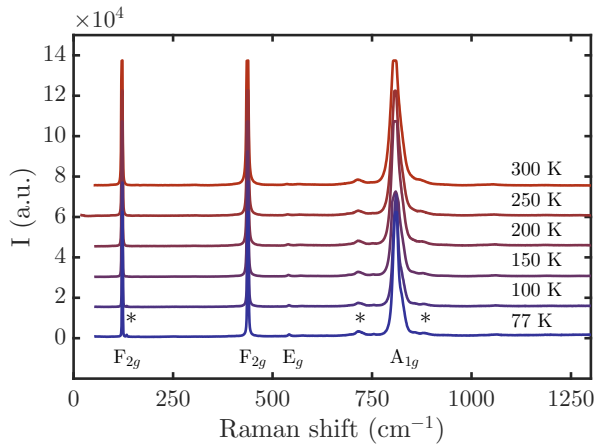}
     	\centering
     	\caption{Temperature dependence of Raman spectra measured on Ba$_{2}$MgWO$_{6}$ single crystals. 4 Raman modes expected for $Fm\bar{3}m$ space group are indicated under the data sets, extra peaks are marked with asterisks at 130, 716 and 880 cm$^{-1}$.} \label{fig:RM_all}
    \end{figure}


    In previously published literature on Ba$_{2}$MgWO$_{6}$ ceramic material by Diao et al. \cite{jr:diel_Ram1}, 3 intense modes and a weak $E_{g}$ mode were identified in Raman spectra collected at 300~K. In our study, we collected Raman spectra from 300~K down to 77~K on a Ba$_{2}$MgWO$_{6}$ single crystal, see Fig.~\ref{fig:RM_all}. The data show some notable differences. The improved signal to noise ratio in our data allow observation of more subtle peaks in the spectra. For all temperatures, three strong and four weak modes are present. When compared to the literature (Tab.~\ref{table:Ram_modes}), we can identify four of these as four expected modes - $F_{2g}$(1), $F_{2g}$(2), $A_{g}$, weak $E_{g}$ - and three remaining are not explained. Three additional peaks have Raman shift of 130, 716 and 880~cm$^{-1}$ (values were obtained from the fit of 77~K data set) and are marked with asterisks in Fig.~\ref{fig:RM_all}. The additional mode at 130~cm$^{-1}$ increases in amplitude upon cooling.

    \begin{table}[h]
    \scriptsize
        \caption{First column represents experimentally defined Raman modes at 300~K in this study (ts). Following columns show the comparison with experimental data at 300~K on the ceramic material~\cite{jr:diel_Ram1}, and two theoretical calculations of phonon spectra~~\cite{jr:diel_Ram1,jr:LDC}.}
        \centering
        \begin{tabular}{l c c c c}
        \hline\\
         & \multicolumn{4}{c}{$\nu$(cm$^{-1}$)}\\
        R. mode & exp(ts) & exp\cite{jr:diel_Ram1} & DFT \cite{jr:diel_Ram1} & LDC\cite{jr:LDC}\\
        \hline \\ 
        F$_{2g}$(1) & 123 & 126 & 105 & 125\\
        F$_{2g}$(2) & 439 & 441 & 405 & 432\\
        E$_{g}$(1) & 536 & 538 & 614 & 522\\
        A$_{1g}$(1) & 807 & 812 & 833 & 776\\
        \hline
        \end{tabular}
        \label{table:Ram_modes}
    \end{table}

\begin{figure}[h]
     	\includegraphics[width=0.9\columnwidth]{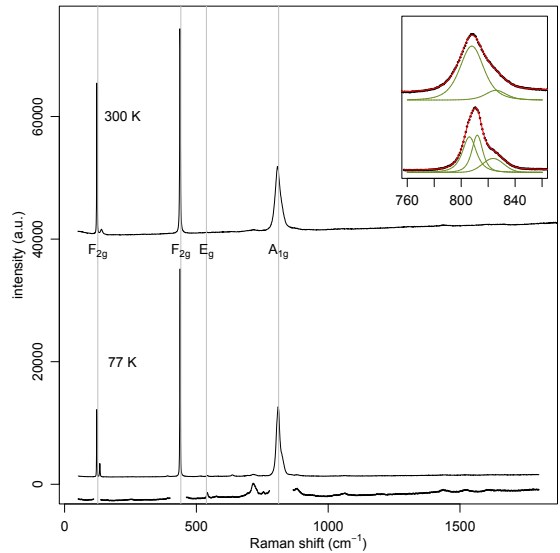}
     	\centering
     	\caption{Raman spectra of Ba$_{2}$MgWO$_{6}$ single crystal at 300 and 77~K. Grey vertical lines mark the position of the 4 Raman active modes found by Diao et al.~\cite{jr:diel_Ram1}. Data depict splitting of lower $F_{2g}$(1) mode, weak $E_{g}$ mode and asymmetric peak of $A_{1g}$ Raman shift. The inset shows close up on $A_{1g}$ mode that can not be characterised by a single Pearson function, even at room temperature. Two components are necessary for data set at 300~K and three for 77~K data. Red line is the fit including a smooth baseline, green the individual components.} 
     	\label{fig:RM_77_300}
    \end{figure}

The $A_{1g}$ mode could not be fit by a single Pearson VII line shape due to the peak's tail at higher frequencies. The width of this mode is reduced with decreasing temperature, uncovering a new vibration very close to the 812~cm$^{-1}$ Raman shift established in \cite{jr:diel_Ram1}. Two peaks at 300~K and three peaks with Pearson VII line shape at 77~K were needed to fit its shape and are plot in the inset of Fig.~\ref{fig:RM_77_300}.

Raman scattering has advantages over diffraction techniques, mainly because of its sensitivity to detect local disorder and symmetry lowering. From the many possible reasons why additional bands may appear in Raman spectra, some are easily eliminated due to the character of Ba$_{2}$MgWO$_{6}$ compound. The electronic configuration of  W$^{6+}$ is 5$d^{0}$, therefore Ba$_{2}$MgWO$_{6}$ is non-magnetic and should have no contribution from magnons. The structural analysis confirmed that there is no Mg/W site mixing. Therefore, the most likely scenario of additional bands comes from local symmetry lowering, which might be due to positional disorder or defects. Slight departure from proposed $Fm\bar{3}m$ space group (on average or locally) can alter the operations of symmetry and the final irreducible representation.

From the nuclear site group analysis (see Table~\ref{table:phonons}), the $A_{1g}$ Raman active mode represents phonon contributions of oxygen atoms only, describing breathing-type vibrations of the octahedra. The extra modes appearing around $A_{1g}$ might be coming from local symmetry lowering. Several reported results in literature \cite{jr:B_site1,jr:B_site2} points to the fact that the $A_{1g}$ mode is sensitive to B-site local disorder in partially ordered perovskites.

    \begin{table}[h]
   \scriptsize
        \caption{Phonon contribution at $\Gamma$ point, nuclear site group analysis for ordered Ba$_{2}$MgWO$_{6}$ double perovskites cubic lattice desribed by Rodrigues et al. \cite{jr:LDC}. $A_{1g}$ and $E_{g}$ Raman modes correspond to oxygen vibrations only.}
        \centering
        \begin{tabular}{l l l}
        \hline\\
        Atom & Site & Phonon contribution \\
        \hline \\ 
        Ba & 8$c$ & F$_{2g}$+F$_{1u}$\\
        Mg & 4$a$ & F$_{1u}$\\
        W & 4$b$ & F$_{1u}$\\
        O & 24$e$ & A$_{1g}$+E$_{g}$\\
        ~ & ~ & +F$_{2u}$+F$_{2g}$+2F$_{1u}$+F$_{1g}$\\
        \hline
        \end{tabular}
        \label{table:phonons}
    \end{table}

\section{Conclusion}

Single crystals of the cubic double perovskite Ba$_{2}$MgWO$_{6}$ have been grown by a new method using BaCl$_{2}$ and MgCl$_{2}$ flux. Obtained crystals have a the average size of 0.4~mm with largest crystals up to 0.6~mm, thus providing an opportunity for material property investigations on single crystals. Heat capacity measurements reveal no phase transitions from 2~K to 300~K for this compound. The temperature dependence of Raman spectra revealed extra modes that are not consistent with 4 proposed Raman active modes established for $Fm\bar{3}m$ space group. For better understanding of Ba$_{2}$MgWO$_{6}$ Raman spectra, more measurements are needed, especially, at $T<$~77~K, to clarify the true symmetry of the compound.

\section*{Acknowledgment}

This work was funded by the European Research Council (ERC) under the European Union’s Horizon 2020 research and innovation program projects HERO (Grant No. 810451). We thank the Interdisciplinary Centre for Electron Miscoscopy at EPFL (CIME), K. Kr\"amer and Daniel Biner at University of Bern for initial growths, and Jens Kreisel for insightful discussions about Raman spectra.

\bibliographystyle{unsrt}

 \bibliography{journals}
    
\end{document}